\documentstyle[editedbook,epsfig,psfig,epsf]{mq}

\def\cm2{cm$^2$ }
\def\se1{s$^{-1}$ }

\begin{opening}
\title{GRBs in the Cannonball model: an overview}
\author{A. De R\'ujula$^1$}
\institute{$^1$ CERN, 1211 Geneva 23}
\end{opening}

\runningtitle{The CB model}
\runningauthor{De R\'ujula}

\begin{document}
\vspace{-0.5cm}
\begin{abstract}
{\small The cannonball model of GRBs is very overt (and, thus, falsifiable)
in its hypothesis and results:
all the considerations I review are based on explicit analytical
expressions derived, in fair approximations, from first principles.
The model provides a good description of
{\it all} the data on {\it all} GRBs of known redshift, 
has made correct predictions, and is unprecedentedly
self-consistent, simple and successful.}
\end{abstract}

\section{Rationale}
The cannonball (CB) model of GRBs \cite{super,DD2000b,optical,radio}
is based on {\bf our} ignorance, for {\bf we} (its authors) do not understand, 
e.g.: how the GRB engine works, 
how core-collapse supernovae (SNe) eject their ejecta,
the transport of angular momentum in processes of
collapse and/or accretion,
relativistic magnetohydrodynamics, the
relativistic ejections in quasars and microquasars...
Thus, we base our starting hypothesis on analogy with the
{\it observations} of quasars and $\mu$-quasars, 
overlooking current numerical simulations of these phenomena\footnote{The
definition \#1.a of ``simulation'' in the OED is: ``The action or practice of 
simulating, with intent to deceive; false pretence, deceitful profession''.}. 
Quasars and $\mu$-quasars
 appear to expel relativistic plasmoids when matter accretes abruptly
from a disk or torus orbiting them.  
We assume the GRB engine to be similar: relativistic
CBs are emitted axially from the recently made compact object in a 
core-collapse SN, as matter that has not been expelled as a 
SN shell (SNS) falls back \cite{yo} to constitute an unstable disk.  
Most indications are that the plasmoids
are made of ordinary matter, not some fancier substance such as 
$e^+\,e^-$ pairs with some finely-tuned ``baryon-load'', as assumed
in the conventional GRB scenarios: fireballs or their progeny 
(hereinafter ``the standard model (SM)''; for a balanced 
review, see \cite{Ghis1}). 

\section{The GRB proper}
\begin{figure}[htb]
\vspace{-1cm}
\centering
\psfig{file=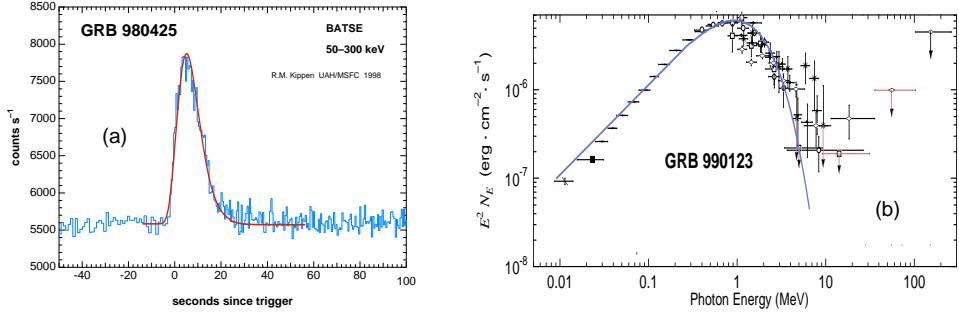,width=13cm}
\vspace{-0.3cm}
\caption{(a) Temporal shape of GRB 980425. (b) $E^2\,dn_\gamma/dE$ 
spectrum of GRB 990123. }
\vspace{-0.2cm}
\label{fig:1}
\end{figure}

Crossing the SNS with a large Lorentz factor $\gamma$, the
surface of a CB is collisionally heated to keV temperatures and the
radiation it emits when it reaches the transparent outskirts of the
shell ---boosted and collimated by the CB's motion---
is a single $\gamma$-ray pulse in a GRB. The cadence
of pulses reflects the chaotic accretion processes and is not 
predictable, but the individual-pulse temporal and spectral properties are.
One example of $\gamma$-ray light curve
is given in Figure \ref{fig:1}a, for the single
pulse of GRB 980425, the closest-by GRB of known redshift $z$.
An example of spectrum is given in Figure \ref{fig:1}b, for the most energetic
recorded GRB of known $z$. The high-energy tail is not well reproduced
by our simplified quasi-thermal model \cite{DD2000b}; it should be flatter.
This is not surprising: we did not take into account that a CB in its rest 
system is bombarded by SNS particles of high $\gamma$, which cannot
be instantaneously thermalized. The low-energy part of the spectrum, in this
and other GRBs, behaves as $E^2\,dn_\gamma/dE\approx E^1$, in agreement
with the CB-model's prediction (the standard fireball scenario inescapably
predicts a slope disagreeing with observation by $\sim 1/2$ unit \cite{GCL2000}).
A long list of general properties of GRB pulses (e.g. that they are narrower at
high than at low energy) is reproduced in the CB-model, in which, unlike
in the SM, the GRBs' $\gamma$'s have a thermal (as opposed
to synchrotron) origin \cite{DD2000b}.

From our analysis of GRBs we deduced that the observed
$\gamma$-ray fluences  
and individual-$\gamma$ energies imply that CBs have typical Lorentz
factors $\gamma\sim 10^3$ and are only observable for angles $\theta$
(between the jet axis and the observer) of ${\cal{O}}(10^{-3})$. For such a small viewing angle, the universal rate of
GRBs and that of core-collapse SNe are comparable: {\it we are defending
the extreme view that a good fraction of such SNe emit GRBs}.
The mass and baryon number ($N_{CB}$)
of a CB are typically a fraction of those of our planet: peanuts,
by stellar standards.

\section{Opening vs. viewing angles}
\begin{figure}[htb]
\vspace{-0.1cm}
\centering
\psfig{file=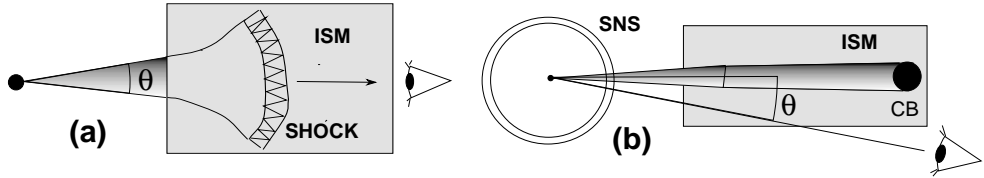,width=13cm}
\vspace{-0.3cm}
\caption{(a)  Standard-Model geometry. (b) CB-model geometry. }
\label{fig:2}
\end{figure}

In GRS 1915+105 the observations are compatible
with the ejecta expanding laterally with a transverse velocity (in their
rest system) comparable to $c/\sqrt{3}$. In many quasars, such as Pictor A,
the ejecta appear to travel long distances without expanding laterally.
In the analysis of the radiation from these sources, as in the CB model
which they inspire, $\gamma$ and $\theta$ (plus the total energy in the 
ejecta) are the parameters needed to describe the
 observations. In the old fireball model of GRBs the ejecta were
spherical. Dar and collaborators have insisted for
a long time that this implied too large a total energy, and that GRBs
should be ``jetted'' emissions from SNe \cite{SD95}. Fireball advocates have
slowly let fireballs become firecones \cite{Rh99} or, more properly, firetrumpets:
jets of material funneled in a cone, with an initial opening angle (also called
$\theta$), that increases as the ejecta encounter the interstellar medium
(ISM), see Figure \ref{fig:2}a.
For years the modellers placed the observer precisely on-axis,
so that all detected GRBs would point to us: an {\it anthropoaxial}
 view. More recently, the SM view is
evolving towards the realization that the observing angle {\it also}
matters \cite{angle1,angle2,angle3,angle4}, a step in what I believe to be the right direction:
the observation angle is the {\it only} one that matters.

We assume CBs, like the observed ejecta in quasars and $\mu$-quasars, 
to contain a tangled magnetic field. In that case, as they plow through the ISM,
they gather and rescatter its constituent protons. The re-emitted protons
exert an inwards pressure on a CB that counters its expansion.
In the approximation of isotropic re-emission in the CB's rest frame
and constant ISM density $n_p$, we explicitly find that, 
in a matter of observer's
minutes, a CB ---faithful to its name--- reaches an asymptotic radius $R$
of ${\cal{O}}(10^{14})$ cm.
In the same approximation we may compute the magnetic field that sustains 
the inwards pressure of the outgoing protons ($B\sim{\cal{O}}$(a few) Gauss)
and derive the explicit law of CB deceleration in the ISM, which depends on
the initial $\gamma=\gamma_0$ as they exit the SNS, and on a ``deceleration''
parameter $x_\infty=N_{CB}/(\pi\,R^2\,n_p)$. CBs decelerate to 
$\gamma (t)=\gamma_0 / 2$ in a journey of $x_\infty/\gamma_0$ length,
typically of ${\cal{O}}(1)$ kpc.

\section{GRB afterglows}
\begin{figure}[htb]
\vspace{-0.1cm}
\centering
\psfig{file=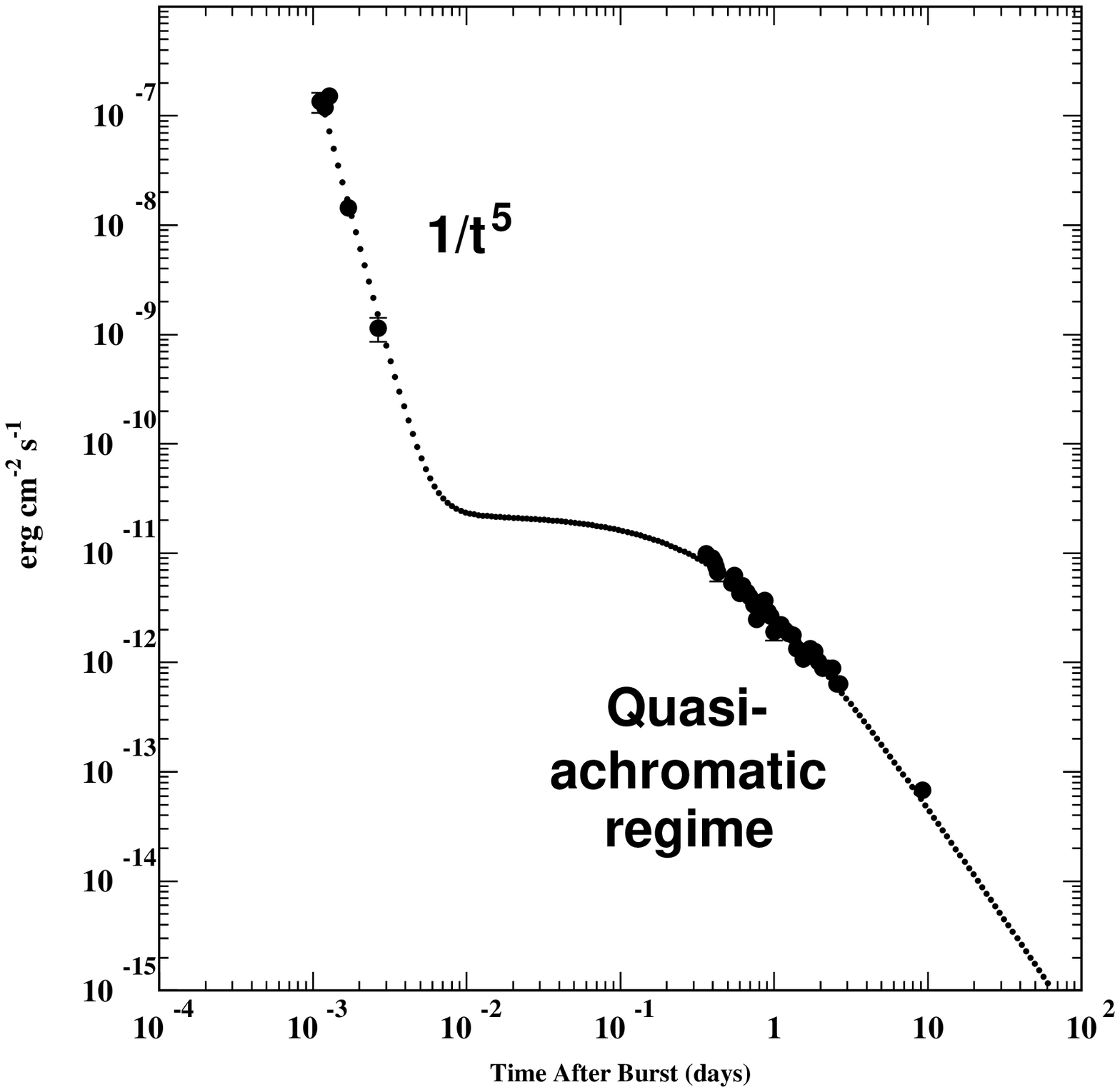,width=6.6cm}\psfig{file=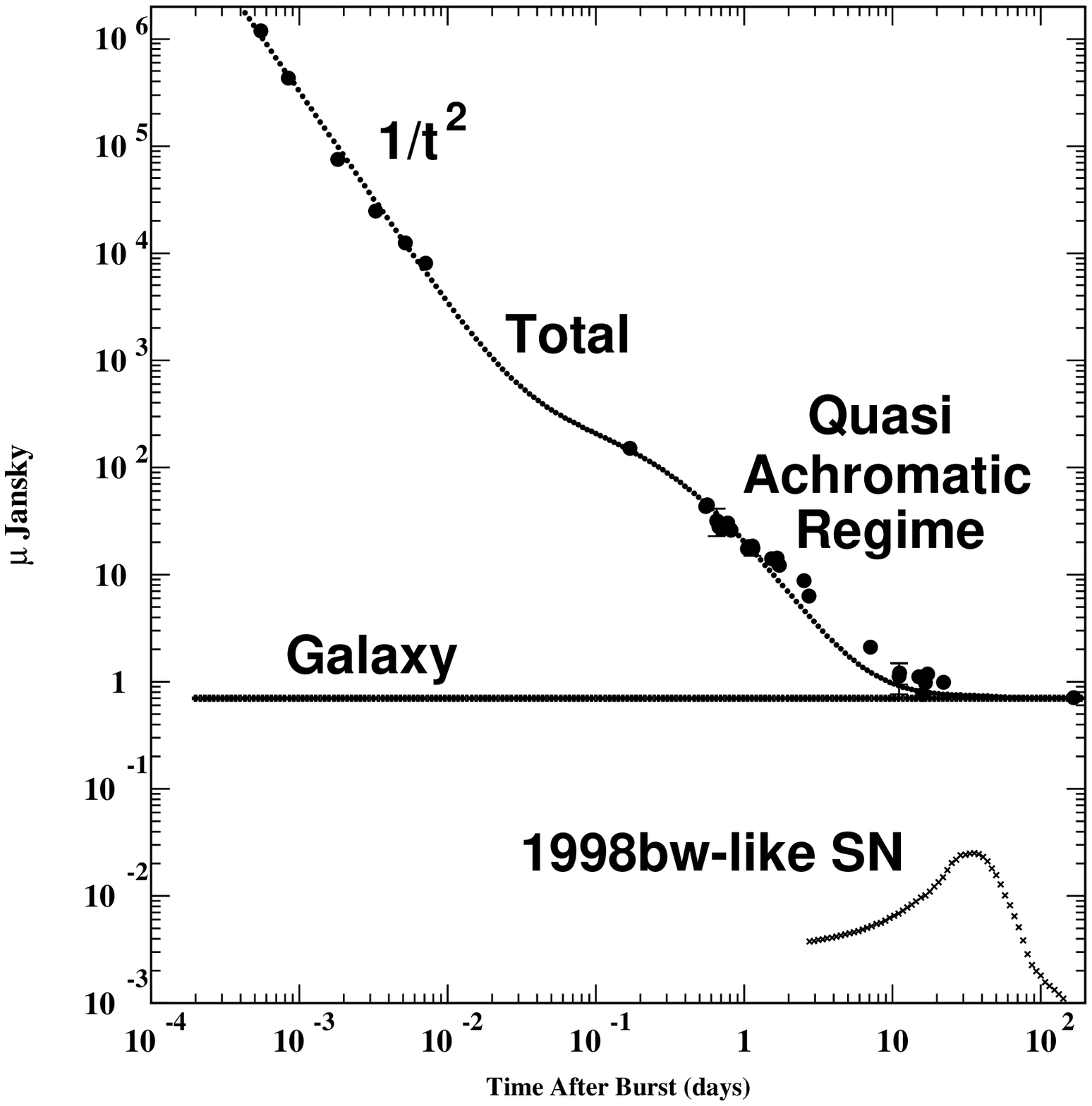,width=6.6cm}
\vspace{-0.2cm}
\caption{(a) X-ray AG of GRB 010222. (b) R-band AG of GRB 990123. }
\vspace{-0.2cm}
\label{fig:3}
\end{figure}

A CB exiting a SNS soon becomes transparent to its own enclosed
radiation. At that point, it is still expanding and cooling adiabatically
and by bremsstrahlung. The bremss spectrum is hard and dominates
the early X-ray AG, with a fluence of predictable magnitude decreasing
with time as $t^{-5}$. An example of how well this describes early X-ray
AGs is shown in Figure \ref{fig:3}a. All  X-ray AGs are
compatible in magnitude and shape with this prediction. 
In the ``internal--external shock'' SM both the
GRB proper and its AG are due to synchrotron radiation,
internal collisions between  shells
result in the GRB,  the external collision of all shells with
the ISM begets the AG. Internal shocks are 
inefficient at creating  internal energy, e.g. two
shells of mass $m$ and Lorentz factors $\gamma$ and
$\gamma/2$ coalesce to produce an object of mass
$\sim 2\,m\,(1+1/16)$ $\sim 1.06\, (2\,m)$, so that, in $e$--$p$--$B$ energy
equipartition,  $\sim 2$\%
of the energy would end up in synchrotron-radiating electrons.
The external shock (a collision
of a composite object of mass $M$ and Lorentz factor $\Gamma$
with the ISM at rest) is overwhelmingly more efficient: a third of all the energy 
$M\,\Gamma\,c^2$ is available! It is difficult to understand 
why the integrated energy in a GRB is much 
{\it larger} than in the AG \cite{Ghis1}, why the X-ray light curves initially
descend  the way they do, and
how the discovery of such patently misbehaving AGs could be hailed
as a great success of the SM (the real difficulty lies in
imagining {\it any} GRB model {\it without} an AG; try!).

The optical AGs of {\it all} GRBs of known $z$  are also well described
in the CB model. They, and the late X-ray AGs,
 are due to synchrotron radiation by the electrons
that the CB gathers in its voyage through the ISM.
The optical AGs for which the data start very early after the GRB
are particularly interesting. The AG of GRB 990123, in Figure \ref{fig:3}b, is
an example. In these early AGs we detect ---in the CB model,
in which the observer's ``clock'' runs at $10^{-6}$ to $10^{-5}$ the rate
of a CB's travel-time--- the CBs plowing through the $\sim r^{-2}$
density-profile of the ``wind'' ejected by the associated SN. This implies an
early decline $\propto t^{-2}$, whose normalization
can also be estimated, also agreeing with the data.
In the SM the absence of ``windy'' signatures is a problem, to the extreme that,
after quoting some 20 earlier failures: 
``Unfortunately, until now there has 
been no clear evidence for a wind-fed circumburst medium'' (CBM), these
SM-devotee authors  \cite{delight} thus continue to report their 
personal feelings about GRB 011121: ``to our delight
[we] have found a good case for a wind-fed CBM''; 
see \cite{Dermer}  for comment.

\section{The GRB/SN association}
\begin{figure}[htb]
\vspace{-0.1cm}
\centering
\psfig{file=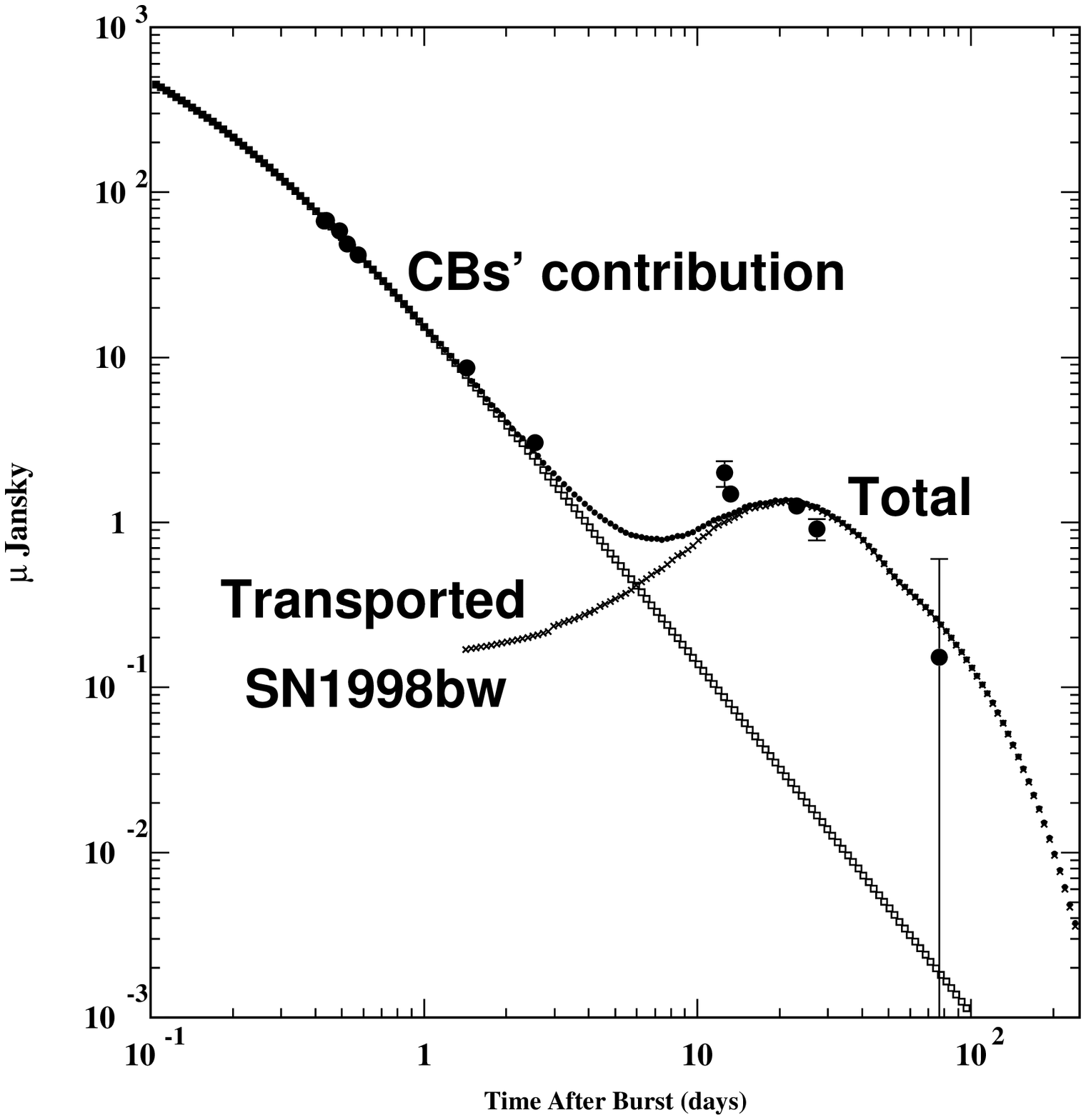,width=7cm}\psfig{file=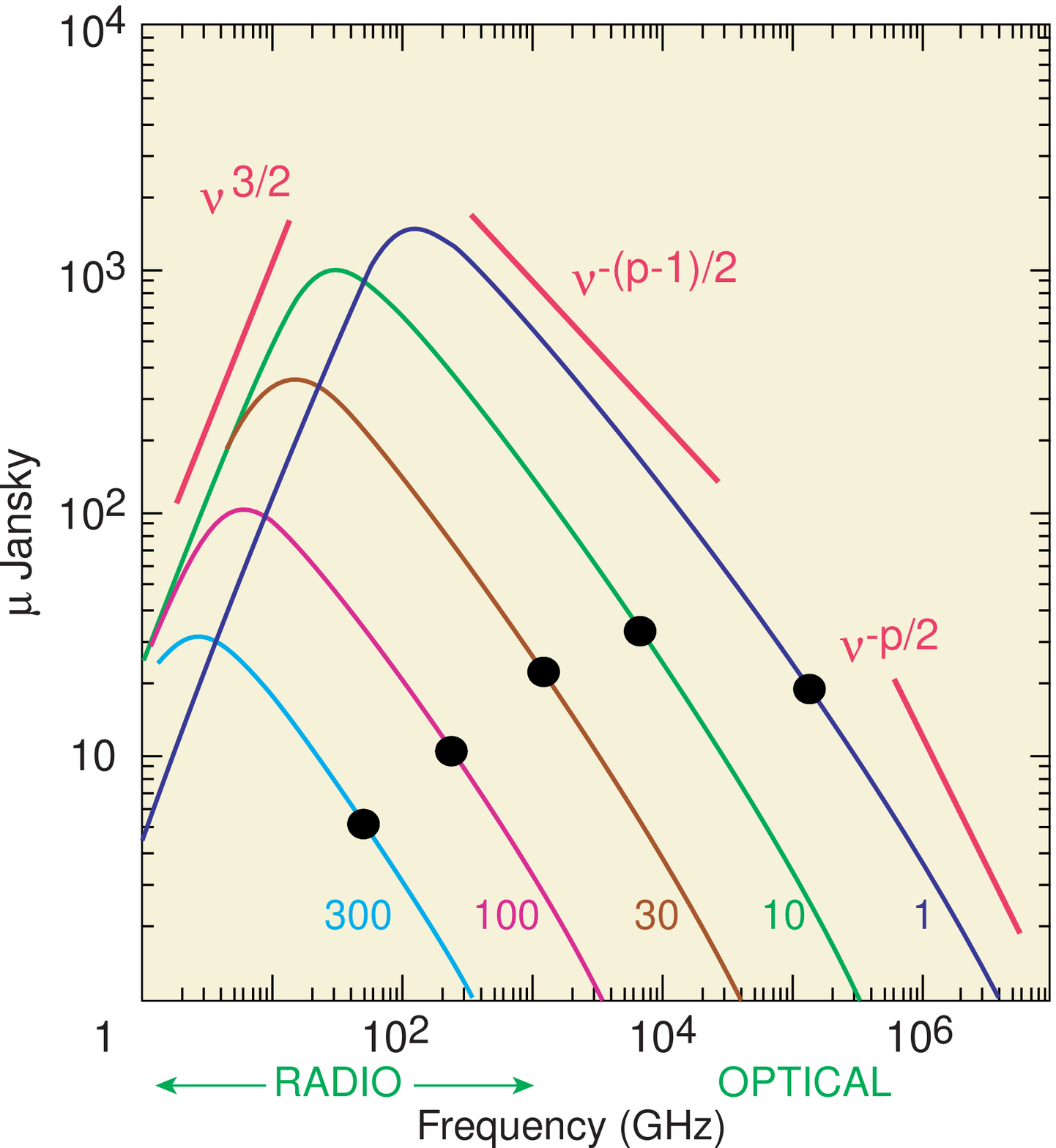,width=6cm}
\vspace{-0.3cm}
\caption{(a) The R-band AG of GRB 011121 with the host-galaxy's 
contribution subtracted. (b) Typical predictions for the CB model's GRB 
spectra, at  times from 1 to 300 days.
The peak frequencies correspond to CB self-opacities of $\cal{O}$(1).
The black dots are the location of the injection-bend frequency
in the synchrotron radiation. }
\vspace{-0.2cm}
\label{fig:4}
\end{figure}

In the CB model, all long-duration GRBs are associated
with SNe compatible with a (properly transported) SN 1998bw \cite{Darbw},
the one associated with the very close-by GRB 980425.
Naturally, ``standard candles'' do not exist, but this one, so far,
is doing a  good job. Half of the score of
GRBs of known $z$ are too far to see their
associated SN and, {\it when the SN should not be seen, it was not seen}. 
Analysed in the CB model, the other half have either 
indications or (as $z$ decreases) incontrovertible evidence for such a SN: 
{\it when the SN could be seen, it was seen}. This gave us confidence
to {\it predict} how the associated SN would appear in the case of GRB 011121.
We used in \cite{pred} the first 2 days of
R-band data to fit the
parameters describing the CBs' contribution to the AG. 
Extrapolating it
in time, we predicted explicitly how the AG would evolve, and we concluded:
{\it the SN will tower in all bands over the CB's declining light curve  
 at day $\sim 30$ after burst}. The comparison with the data \cite{121},
gathered later, is shown in Figure \ref{fig:4}a.
The SN spectrum is slightly
bluer than that of 1998bw, but not significantly so.
In the SM, it is not possible to reproduce the previous exercise
because the AGs (unlike the smoothly-varying
data) have ``breaks''. The early data on GRB 011121
do not tell you where the break ``is'': they cannot be extrapolated. That may 
help explain why the same SM-abiding observers first
concluded that this GRB had no associated SN \cite{1}, the day after that it did 
\cite{2}, to
compromise finally into half of a SN1998bw-like signal \cite{delight}. 

\section{The spectra of GRB afterglows}
\subsection{The injection bend}
\begin{figure}[htb]
\vspace{-0.1cm}
\centering
\psfig{file=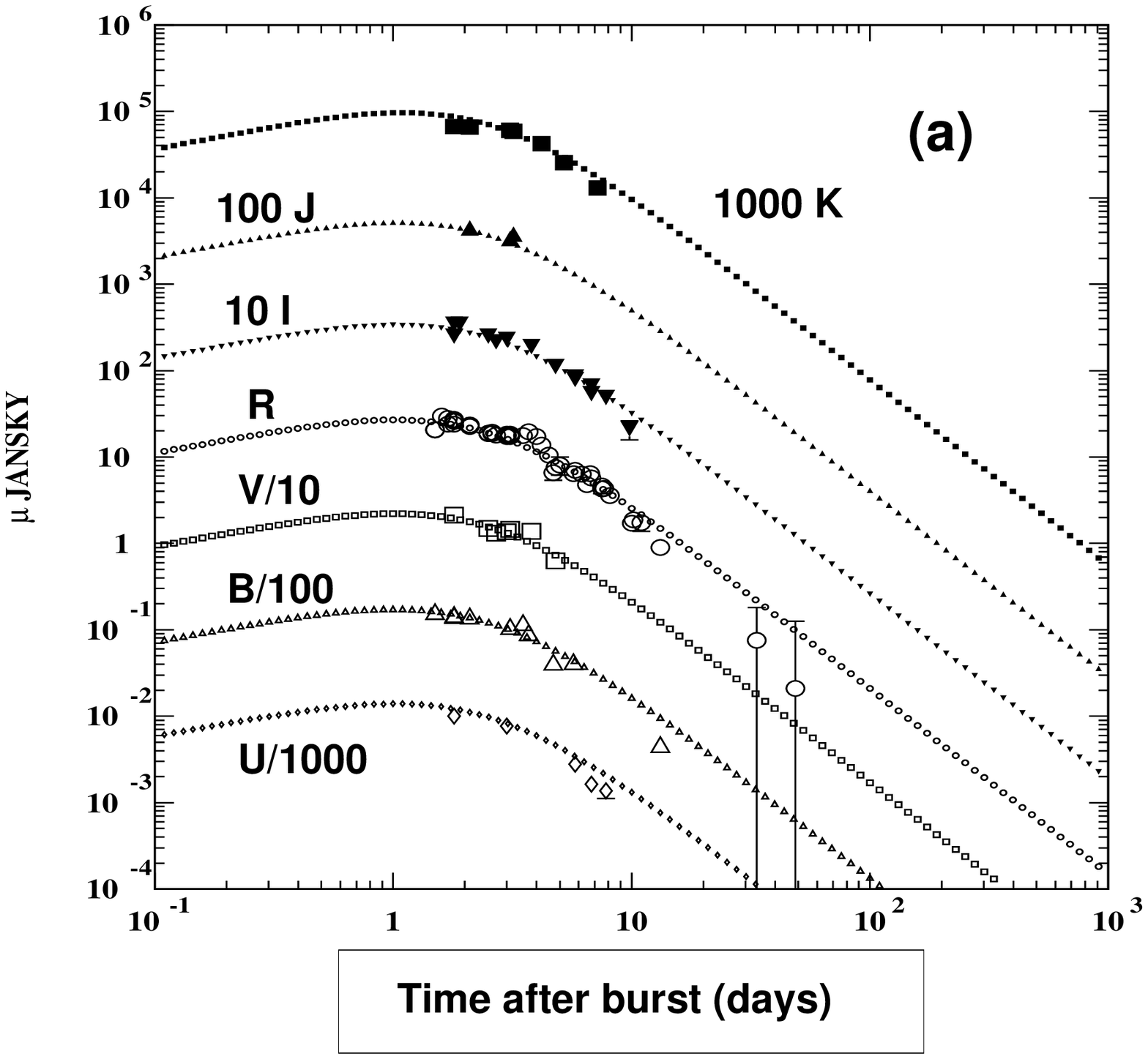,width=6cm}\psfig{file=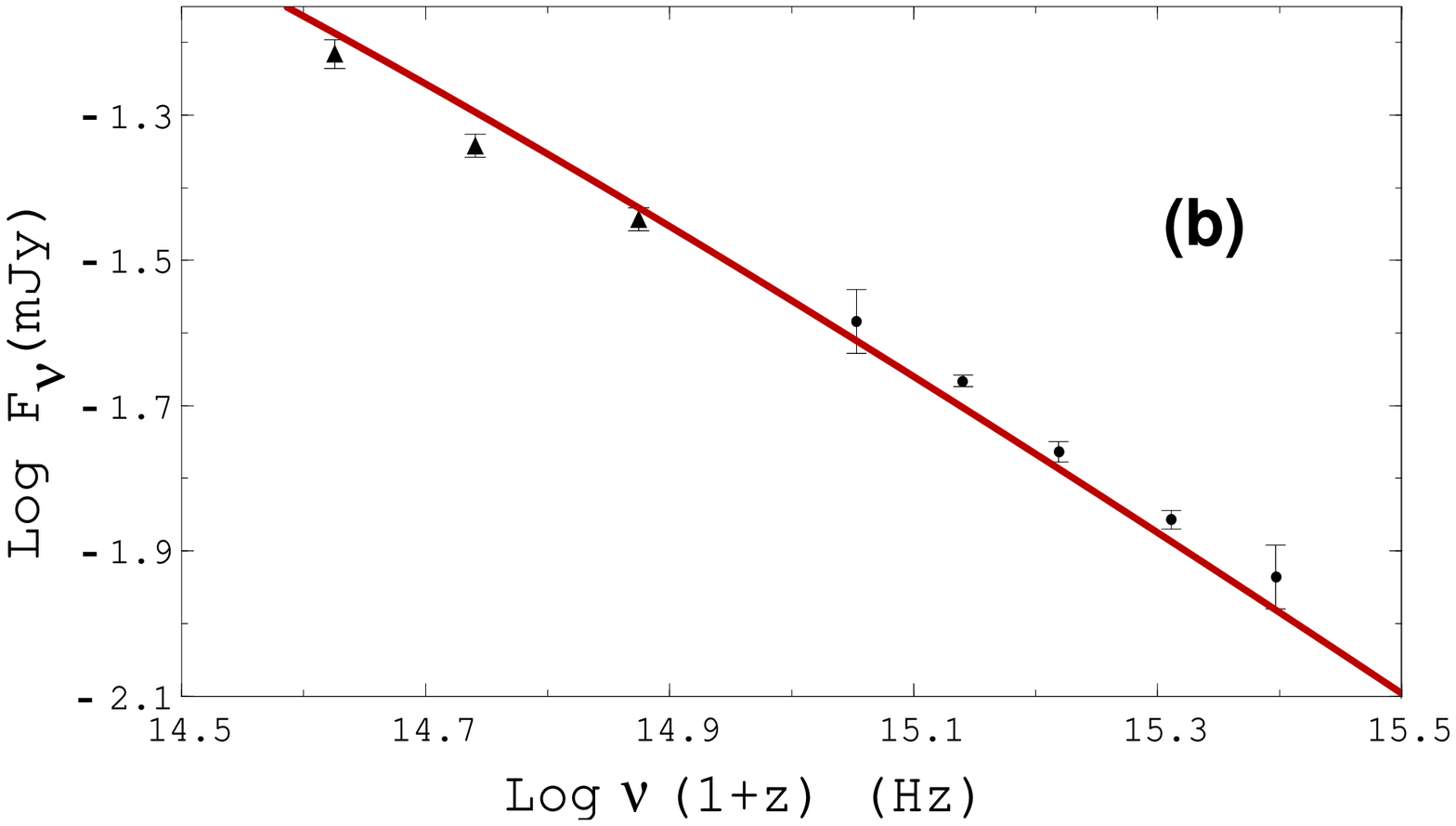,width=7.5cm}
\vspace{-0.2cm}
\caption{GRB 000301c. (a) Host-galaxy-subtracted
CB fit to optical light curves in all bands. (b) Predicted spectrum at 
$t\sim 3$ days, in an optical window. }
\vspace{-0.2cm}
\label{fig:5}
\end{figure}

In the CB model  \cite{radio}   the spectrum of electrons in the
CB, accelerated by its enclosed magnetic maze and cooled by synchrotron
radiation, has an {\it injection bend} at the energy $E_b=m_e\,c^2\,\gamma(t)$
at which a CB would, in its rest system, see the ISM electrons arrive,
with $\gamma(t)$ the instantaneous CB Lorentz factor, extractable from
the fit to the AG light curve. The emitted synchrotron radiation has
a corresponding bend at a frequency (in the CB's frame) 
$\nu_b=0.2175\,\gamma(t)^2\,\nu_L$, with $\nu_L$ the Larmor frequency
in the CB's magnetic field, whose magnitude is also explicit and evolves
as $B(t)\propto \gamma(t)$, so that $\nu_b\propto \gamma(t)^3$. 
Prior to absorption corrections, the synchrotron fluence in a CB's 
rest system is:
\begin{equation}
\vspace{-.2cm}
f_{sync}\equiv\nu\,{dn_\gamma/ d\,\nu} \propto
[\nu/\nu_b]^{-1/2}\;
\left(1+[\nu/\nu_b]^{(p-1)}\right)^{-1/2}
\label{sync}
\end{equation}
with a predicted $p\!\approx\! 2.2$, in agreement with all the data on
relatively late optical AGs, at which time $\nu\!\gg\! \nu_b$ and
$f_{sync}\!\propto\! \nu^{-\alpha}$ with $\alpha\!=\!p/2\!\approx\! 1.1$
(it is easier to extract this index from fits to the AG light curve than
from the spectra, which are beset by absorption corrections).
The explicit interpolating form of Eq.~\ref{sync} is a guess, but
the existence and explicit time-dependence of the injection bend are 
bold conclusions, to be confronted with data.
Getting a bit ahead of myself, I show in Figure \ref{fig:4}b a 
typical predicted  spectrum, in the observer's frame.
The figure shows how the predicted frequency of the spectral
bend  diminishes with time. Measured around a fixed
frequency, a spectral slope may be time-dependent: 
$\alpha\!=\! (p-1)/2\!\approx\! 0.6$ before the ``passage'' of the injection
bend and $\alpha\! =\! p/2\!\approx\! 1.1$ after it.
 In the 7 cases of GRBs with sufficient
data to do this test, the agreement with expectations is good
(the errors are often large). The complementary
test is to look at a narrow-band spectrum  when the bend is
crossing or is nearby, so that the predicted slope would be neither
of the extremes. A good case, with insignificant absorption,
is GRB 000301c. From its observed optical light curves,
fit to the CB model as in Figure \ref{fig:5}a, we extract the AG parameters
(normalization, $\theta$, $\gamma_0$ and $x_\infty$) needed to
predict $\nu_b(t)$ and the spectral shape at a given time.
The results at $t\sim 3$ days
are shown in Figure \ref{fig:5}b (the normalization is borrowed from
the data, but the slightly curving slope is a {\it 
prediction}).

\subsection{Broad-band spectra}
\begin{figure}[htb]
\vspace{-0.4cm}
\centering
\psfig{file=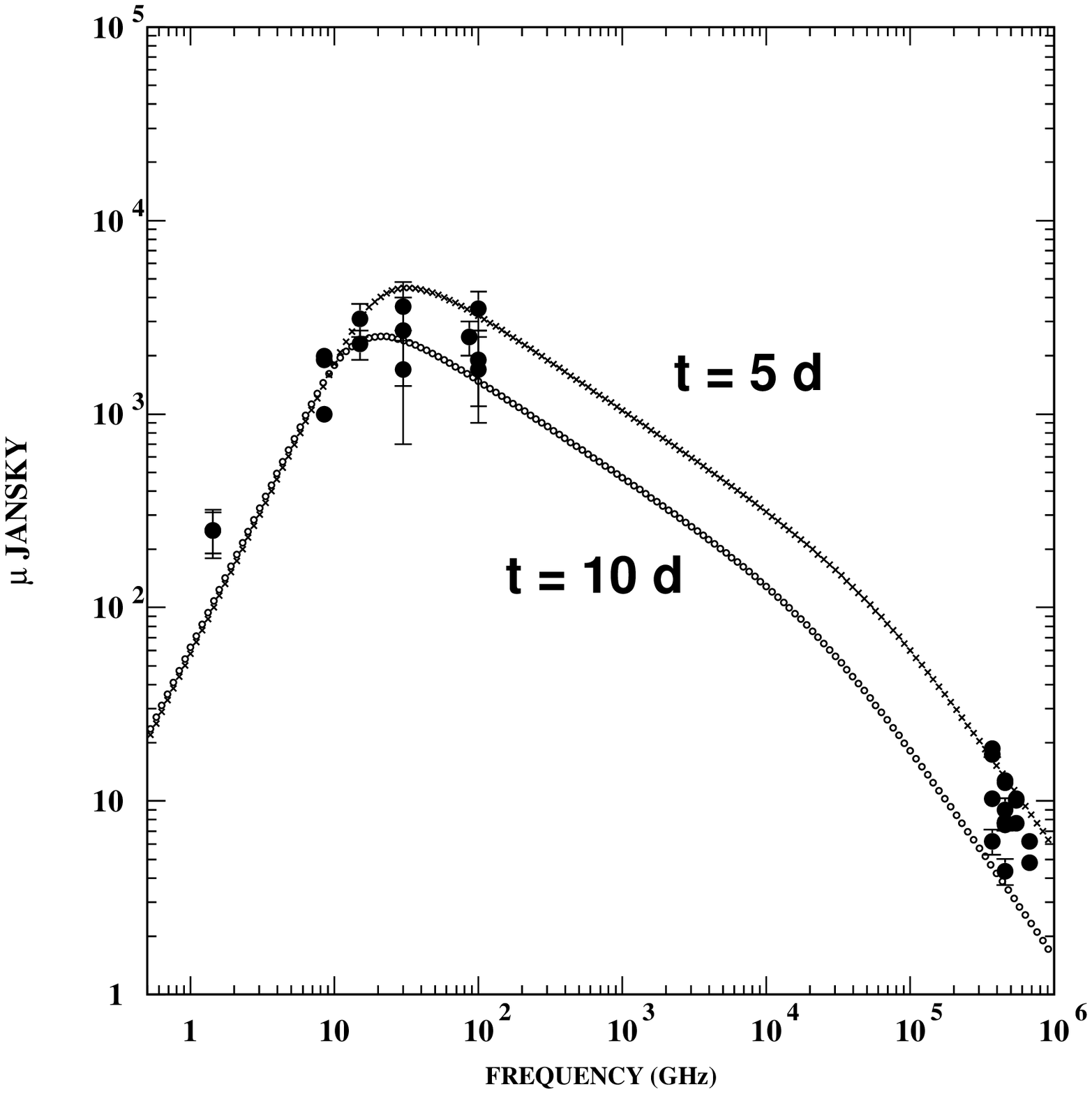,width=6.cm}\psfig{file=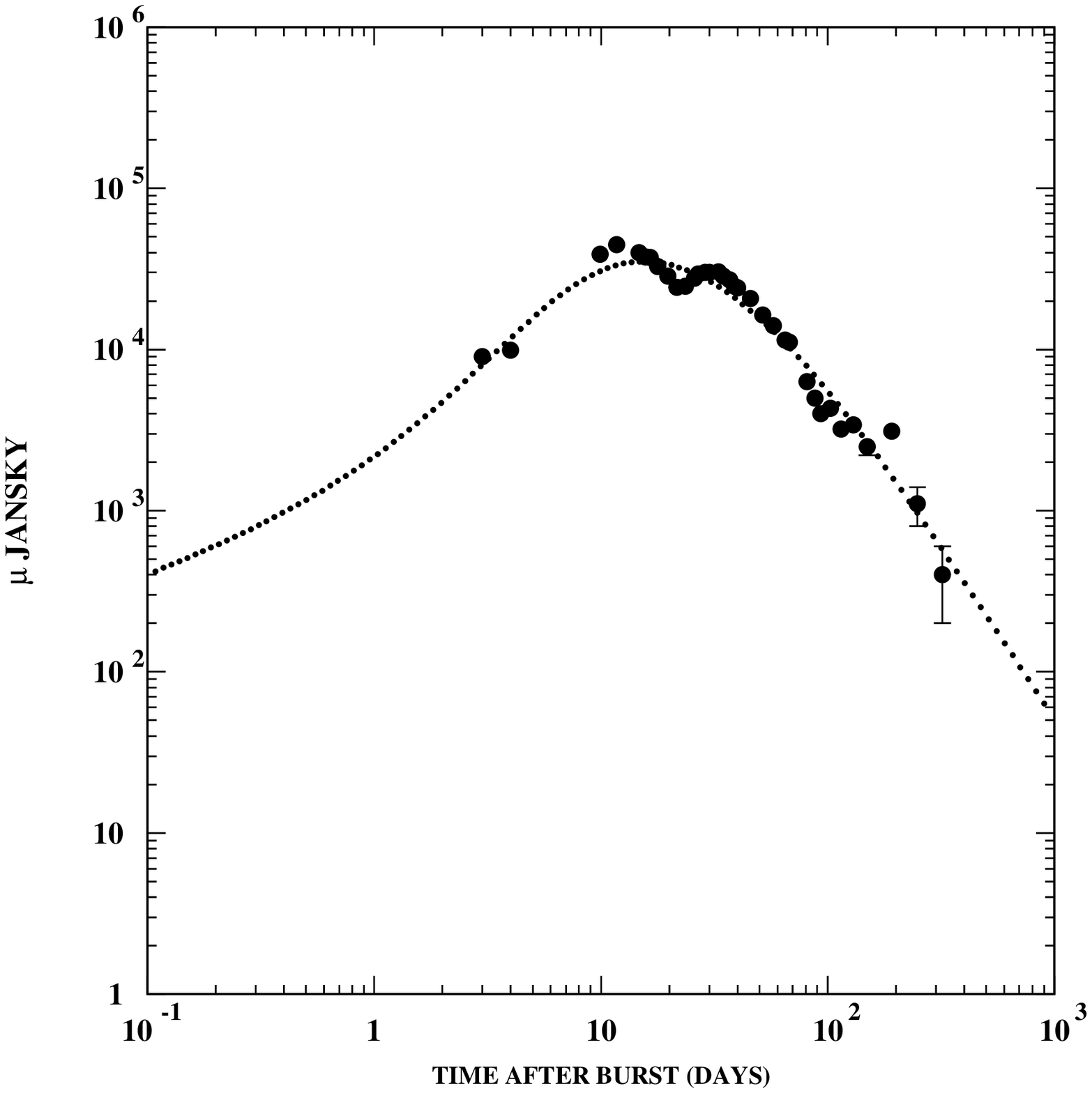,width=6.cm}
\vspace{-0.4cm}
\caption{(a) The CB-model fit to the broad-band spectrum of GRB 991208 at 
$t=5$ to 10 days. (b) 4.8 GHz light curve of  GRB 980425. At other times and
frequencies the fits are equally good.}
\vspace{-0.1cm}
\label{fig:6}
\end{figure}

In the radio domain, self-absorption in the CB is important. The dominant
mechanism is free-free attenuation, characterized by a single parameter
$\nu_a$ in the opacity, which behaves as 
$\tau_\nu=(\nu_a/\nu)^2(\gamma(t)/\gamma_0)^2$. Absorption is responsible
for the turn-around of the spectra in Figure \ref{fig:4}b. All observed
spectra agree well, in spite of the scintillations in the radio, with this figure,
fit in each case to the specific GRB. 
The most complete broad-band data are perhaps those of GRB 991208.
In Figure \ref{fig:6}a I show its spectrum between 5 and 10 days 
after burst.

The overall fits to the light curves
and spectra in all bands involve the four parameters mentioned earlier plus
the {\it single} ``radio'' parameter $\nu_a$ (the situation is in stark contrast with
the SM model fits that are multiple-choice and involve many parameters
in the spectral description, sometimes re-fit for each particular observational
time). The complete description of the radio AG requires the inclusion of
two effects that, in fair approximations, 
introduce no extra parameters:  a ``cumulation factor''
for the electrons that emit the observed radio frequencies
(it takes time for the ISM electrons gathered by the CB to cool to
the radio-emitting energies) and an ``illumination and limb-darkening''
factor taking into account that the CBs are viewed relativistically
(an observer would ``see'' almost all of the $4\pi$ surface of a spherical CB).

\subsection{GRB 980425 and its associated SN: 1998bw}
\begin{figure}[htb]
\vspace{-0.1cm}
\centering
\psfig{file=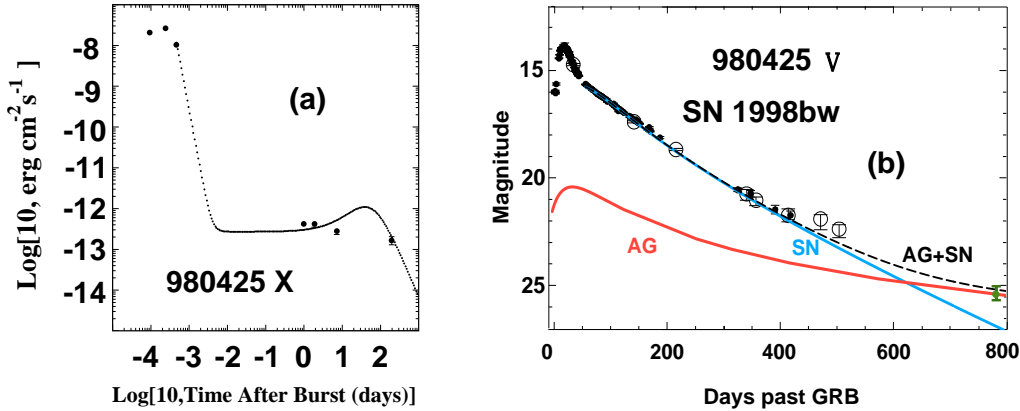,width=13.5cm}
\vspace{-0.3cm}
\caption{CB-model fits to GRB 980425. (a) The X-ray AG. (b) The V-band AG:
the SN contribution, the CB's contribution and the total. All parameters
(but $z$ and $\theta$) are ``normal''. }
\label{fig:7}
\end{figure}

To give an example of radio light curve I choose the most interesting
of all GRBs: 980425, at a tiny redshift of $z=0.0085$; see
 Figure \ref{fig:6}b \cite{radio}.  
In the CB model, this GRB and its associated
SN1998bw are {\bf not} exceptional. Because it was viewed at
an exceptionally large angle ($\sim 8$ mrad), its 
$\gamma$-ray fluence was comparable to that of more distant
GRBs, viewed at  $\theta\sim 1$ mrad. That is why its optical AG
was dominated by the SN, except for the last measured point
\cite{super}. The X-ray AG (Figure \ref{fig:7}a)
of its single CB (see Figure \ref{fig:1}a)
is also of ``normal'' magnitude, it is {\bf not} emitted by the SN,
and its fitted parameters allowed us to predict successfully the 
magnitude of the cited 
last optical point \cite{optical}, see Figure \ref{fig:7}b.
The normalization, time and frequency
dependence of the radio AG of this GRB are also ``normal'', and due to
the CB, {\bf not} the SN \cite{radio}. SN1998bw, deprived of its ``abnormal''
X-ray and radio emissions (which it did not emit!), loses most of
its ``peculiarity''.

\section{Some other predictions}
The analysis of the radio scintillations of pulsars is one
way to measure their sky-projected velocities, in
agreement with proper-motion results.
For cosmological GRBs, the sky angular velocity of their 
CBs happens to be comparable to that of the much
slower and closer-by Galactic pulsars. Perhaps, then,
the analysis of GRB radio oscillations may result in a
 measurement of their apparent velocities, which are
``hyperluminal'': $v_T={\cal{O}}(10^2)\, c$! \cite{radio}.

For lack of time, I have not discussed the mounting evidence for X-ray lines 
in GRB AGs. The results \cite{Xlines2} are quite intriguing:
the alleged lines are at the positions {\it predicted} \cite{Xlines}
in the CB model. In the SM model the observed features
in X-ray spectra are supposed to be Fe lines or recombination edges,
or characteristic lines of a variety of ``metals''. In the CB model
the lines ought to be emitted by the light constituents of a CB (mainly
H and He). These lines are strongly blue-shifted by the CB's
relativistic motion, the corresponding Doppler factor, as a function of
time, can be extracted from the CB-model's parametrization
of the optical or X-ray AG light curves. Thus, for the cases where the 
red-shift is known, the observed line energies are {\it predicted}.
Current data are not precise enough, but in
future observations the time-dependence of these lines may
be observable (the CBs decelerate, and the Doppler factor diminishes with
time).

\section{Avatars and limitations of the CB model}

The abstract summarizes my conclusions: they need not be repeated, but...

What are the CB-model's limitations? We contend that CBs are 
emitted at a time $t_{_{CB}}={\cal{O}}(1)$ day after the
parent-star's core-collapse \cite{yo}. That is the typical time for 
not-expelled and not-imploded stellar
material to collapse back to the newly made compact object.
It is peculiar that the GRBs typically last a small fraction
of $t_{_{CB}}$, but the build-up of an unstable
accretion disk may take long, while its episodes of ``fall'' may be 
brief. With $t_{_{CB}}={\cal{O}}(1)$ day, the SNS has moved to a distance
that plays a role in our good description of the duration
of GRB pulses. Yet, our model of the complex CB--SNS collisions may be
naive: we know that a more detailed model will result in a smaller implied 
$t_{_{CB}}$.  

As we move away from these early violent
collisions into the subsequent AG era, the CB model becomes simpler and
its results and successes are robust. When we move even further and confront 
much of
the GRB community of this planet... that is when the problems hit the roof.

\section*{Aknowledgements} 
I enjoyed discussing with ---and learning from--- many participants at
 this conference,
particularly Alberto Castro-Tirado, Philippe Durouchoux, 
Sophie Ferry, Dick Hubbard, Felix Mirabel and
Enrico Ramirez-Ruiz.


\begin{thebibliography}{}

\bibitem{super}
Dar A. \& De R\'ujula A., astro-ph/0008474 
\bibitem {DD2000b}
Dar A. \& De R\'ujula A.,  astro-ph/0012227 
\bibitem{optical}
Dado, S., Dar, A. \& De R\'ujula, A., 2002, 
 A\&A 388, 1079
\bibitem{radio}
Dado, S., Dar, A. \& De R\'ujula, A.,  submitted to A\&A, 
(astro-ph/0204474)
\bibitem{yo}
De R\'ujula, A., 1987, Phys. Lett. 193, 514
\bibitem{Ghis1}
Ghisellini, G., astro-ph/0111584
\bibitem{GCL2000}
Ghisellini, G. et al., MNRAS, 313, L1
\bibitem{SD95}
Shaviv N.J., \& Dar, A., 1995, ApJ 447, 863
\bibitem{Rh99}
Rhoades J.E., ApJ 525, 737
\bibitem{angle1}
Rossi, E. et al. astro-ph/0112083
\bibitem{angle2} 
Zhang, B. \& Meszaros, P.L., astro-ph/0112118;
\bibitem{angle3}
Salmonson J.D. \& Galama T.J., astro-ph/0112298
\bibitem{angle4}
Granot J. et al., astro-ph/0201322
\bibitem{delight}
Price P.A. et al., astro-ph/0203467
\bibitem{Dermer}
Dermer, C.D., astro-ph/0204037
\bibitem{Darbw}
Dar A., GCN Circ. 346
\bibitem{pred}
Dado, S., Dar, A. \& De R\'ujula, A., 
 astro-ph/0111468
\bibitem{121}
Dado, S., Dar, A. \& De R\'ujula, A.,  2002,
 ApJL  572, L143
\bibitem{1}
Bloom J.S., et al., 2002 GCN Circ. 1274 
\bibitem{2}
Kulkarni S., et al., 2002 GCN Circ. 1276 
\bibitem{Xlines2}
Dado, S., Dar, A. \& De R\'ujula, A.,  submitted to ApJ 
(astro-ph/0207015)
\bibitem {Xlines}
Dar A. \& De R\'ujula A., astro-ph/0102115 

 
 




\end{thebibliography}
\end{document}